\documentclass{article}
\usepackage{spconf,amsmath,epsfig}
\ninept

\title{Adaptive Reduced-Rank Processing Using a Projection Operator Based on Joint
Iterative Optimization of Adaptive Filters For CDMA Interference Suppression }
\name{Rodrigo C. de Lamare $\dagger$ and Raimundo Sampaio-Neto
 $\ddagger$
}
\address{  $\dagger$  Communications Research Group, University of York, United Kingdom \\
 $\ddagger$  CETUC, Pontifical Catholic University of Rio de Janeiro
(PUC-RIO), Brazil \\
E-mails: rcdl500@ohm.york.ac.uk, raimundo@cetuc.puc-rio.br}
\begin{document}
%
\maketitle
\begin{abstract}
This paper proposes a novel adaptive reduced-rank filtering scheme
based on the joint iterative optimization of adaptive filters. The
proposed scheme consists of a joint iterative optimization of a
bank of full-rank adaptive filters that constitutes the projection
matrix and an adaptive reduced-rank filter that operates at the
output of the bank of filters. We describe minimum mean-squared
error (MMSE) expressions for the design of the projection matrix
and the reduced-rank filter and simple least-mean squares (LMS)
adaptive algorithms for its computationally efficient
implementation. Simulation results for a CDMA interference
suppression application reveals that the proposed scheme
significantly outperforms the state-of-the-art reduced-rank
schemes, while requiring a significantly lower computational
complexity.
\end{abstract}
\begin{keywords}
{Adaptive filters, iterative methods.}
\end{keywords}

\section{Introduction}
\label{sec:intro}

In the literature of adaptive filtering \cite{diniz}, the designer
can find a huge number of algorithms with different trade-offs
between performance and complexity. They range from the simple and
low-complexity least-mean squares (LMS) algorithms to the fast
converging though complex recursive least-squares (RLS)
techniques. In the last decades, several attempts to provide
cost-effective adaptive filters with fast convergence performance
have been made through variable step-size algorithms,
data-reusing, sub-band and frequency-domain adaptive filters and
RLS type algorithms with linear complexity. A challenging problem
that remains unsolved by conventional techniques is that when the
number of elements in the filter is large, the algorithm requires
a large number of samples (or data record) to reach its
steady-state behavior. In these situations, even RLS algorithms
require an amount of data proportional to $2M$ \cite{diniz}, where
$M$ is the number of elements of the filter, in order to converge
and this may lead to unacceptable convergence and tracking
performance. Furthermore, in highly dynamic systems such as those
found in wireless communications, large filters usually fail or
provide poor performance in tracking signals embedded in
interference and noise.

An alternative and effective technique in low sample support
situations and in problems with large filters is reduced-rank
parameter estimation \cite{scharf}-\cite{delamaresp}. The
advantages of reduced-rank adaptive filters are their faster
convergence speed and better tracking performance than existing
full-rank techniques when dealing with large number of weights.
Several reduced-rank methods and systems are based on principal
components analysis, in which a computationally expensive
eigen-decomposition is required \cite{bar-ness}-\cite{song&roy} to
extract the signal subspace . Other recent techniques such as the
multistage Wiener filter (MWF) of Goldstein {\it et al.} in
\cite{gold&reed} perform orthogonal decompositions in order to
compute its parameters, leading to very good performance at the
cost of a relatively high complexity and the existence of
numerical problems for implementation.

In this work we propose an adaptive reduced-rank filtering scheme
that employs a projection matrix based on combinations of adaptive
filters. The proposed scheme consists of a joint iterative
optimization of a bank of full-rank adaptive filters that
constitutes the projection matrix and an adaptive reduced-rank
filter that operates at the output of the bank of full-rank
filters. We describe MMSE expressions for the design of the
projection matrix and the reduced-rank filter along with simple
LMS adaptive algorithms for its computationally efficient
implementation. We assess the performance of the proposed scheme
via simulations for CDMA interference suppression.

The rest of this paper is organized as follows. Section 2 states
the basic reduced-rank filtering problem. Section 3 presents the
novel reduced-rank scheme, the joint iterative optimization
approach and the MMSE design of the filters. Section 4 introduces
LMS algorithms for implementing the new scheme. Section 5 presents
and discusses the numerical simulation results, while Section 6
gives the conclusions.

\section{Reduced-Rank MMSE Parameter Estimation and Problem Statement}
\label{sec:format}

The MMSE filter is the parameter vector ${\bf w}= \big[w_1^{}~
w_2^{} ~ \ldots ~ w_M^{}\big]^T$, which is designed to minimize
the MSE cost function
\begin{equation}
J = E\big[ | d(i) - {\bf w}^{H}{\bf r}(i)|^2 \big],
\end{equation}
where $d(i)$ is the desired signal, ${\bf r}(i)=[r_{0}^{(i)}~
\ldots ~r_{M-1}^{(i)}]^{T}$ is the received data, $(\cdot)^{T}$
and $(\cdot)^{H}$ denote transpose and Hermitian transpose,
respectively, and $E[\cdot]$ stands for expected value. The set of
parameters ${\bf w}$ can be estimated via standard stochastic
gradient or least-squares estimation techniques \cite{diniz}.
However, the laws that govern the convergence behavior of these
estimation techniques imply that the convergence speed of these
algorithms is proportional to $M$, the number of elements in the
estimator. Thus, large $M$ implies slow convergence. A
reduced-rank algorithm attempts to circumvent this limitation in
terms of speed of convergence and tracking capabilities by
reducing the number of adaptive coefficients and extracting the
most important features of the processed data. This dimensionality
reduction is accomplished by projecting the received vectors onto
a lower dimensional subspace. Specifically, let us introduce an $M
\times D$ projection matrix ${\bf S}_{D}$ that carries out a
dimensionality reduction on the received data as given by
\begin{equation}
\bar{\bf r}(i) = {\bf S}_D^H {\bf r}(i),
\end{equation}
where, in what follows, all $D$-dimensional quantities are denoted
with a "bar". The resulting projected received vector $\bar{\bf
r}(i)$ is the input to a tapped-delay line filter represented by
the $D \time 1$ vector $\bar{\bf w}=\big[ \bar{w}_1^{}
~\bar{w}_2^{}~\ldots\bar{w}_D^{}\big]^T$ for time interval $i$.
The filter output corresponding to the $i$th time instant is
\begin{equation}
x(i) = \bar{\bf w}^{H}\bar{\bf r}(i).
\end{equation}
If we consider the MMSE design in (3) with the reduced-rank
parameters we obtain
\begin{equation}
\bar{\bf w} = \bar{\bf R}^{-1}\bar{\bf p},
\end{equation}
where $\bar{\bf R} = E[ \bar{\bf r}(i)\bar{\bf r}^{H}(i)]={\bf
S}_D^H{\bf R}{\bf S}_D$ is the reduced-rank covariance matrix,
${\bf R} = E[{\bf r}(i){\bf r}^{H}(i)]$ is the full-rank
covariance matrix, $\bar{\bf p}=E[d^*(i)\bar{\bf r}(i)]={\bf
S}_D^H{\bf p}$ and ${\bf p}=E[d^*(i){\bf r}(i)]$. The associated
MMSE for a rank $D$ estimator is expressed by
\begin{equation}
J = \sigma^2_d - \bar{\bf p}^H \bar{\bf R}^{-1}\bar{\bf p},
\end{equation}
where $\sigma^2_d$ is the variance of $d(i)$. Based upon the
problem statement above, the rationale for reduced-rank schemes
can be simply put as follows. How to efficiently (or optimally)
design a transformation matrix ${\bf S}_D$ with dimension $M
\times D$ that projects the observed data vector ${\bf r}(i)$ with
dimension $M \times 1$ onto a reduced-rank data vector $\bar{\bf
r}(i)$ with dimension $D \times 1$? In the next section we present
a novel approach based on the joint optimization of a projection
matrix formed by a bank of full-rank adaptive filters and a
reduced-rank adaptive filter.

\section{Proposed Reduced-Rank Scheme }
\label{sec:pagestyle}

In this section we detail the principles of the proposed
reduced-rank scheme using a projection operator based on adaptive
filters. The new scheme, depicted in Fig. 1, employs a projection
matrix ${\bf S}_{D}(i)$ with dimension $M \times D$ to process a
data vector with dimension $M \times 1$, that is responsible for
the dimensionality reduction, and a reduced-rank filter $\bar{\bf
w}(i)$ with dimension $D \times 1$, which accomplishes the second
stage of the estimation process over a reduced-rank data vector
$\bar{\bf r}(i)$ to produce a scalar estimate $x(i)$. The
projection matrix ${\bf S}_{D}(i)$ and the reduced-rank filter
$\bar{\bf w}(i)$ are jointly optimized in the proposed scheme
according to the MMSE criterion.

\begin{figure}[htb]
 \vspace*{-2em}
       \centering  
       \hspace*{-4.5em}
       {\includegraphics[width=11.5cm, height=5cm]{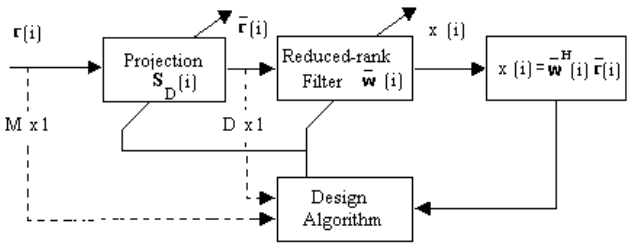}}
       \vspace*{-4em}
       \caption{ Proposed Reduced-Rank Scheme.}

\end{figure}

Specifically, the projection matrix is structured as a bank of $D$
full-rank filters ${\bf s}_d(i)=[s_{1,d}(i) ~ s_{2,d}(i)~
\ldots~s_{M,d}(i) ]^T$ ($d = 1,~\ldots,~D$) with dimension $M
\times 1$ as given by ${\bf S}_{D}(i) = [~{\bf s}_1(i) ~| ~{\bf
s}_2(i)~| ~\ldots~|{\bf s}_D(i)~]$. Let us now mathematically
express the output estimate $x(i)$ of the reduced-rank scheme as a
function of the received data ${\bf r}(i)$, the projection matrix
${\bf S}_D(i)$ and the reduced-rank filter $\bar{\bf w}(i)$ (we
will drop index (i) of the components for ease of presentation):
\begin{equation}
\begin{split}
x(i) & = (\bar{w}_1^*s_{1,1}^*r_1 + \bar{w}_1^*s_{2,1}^*r_2 +
\ldots \bar{w}_1^*s_{M,1}^*r_M) + \\ & \quad
(\bar{w}_2^*s_{1,2}^*r_1 + \bar{w}_2^*s_{2,2}^*r_2 + \ldots
\bar{w}_2^*s_{M,2}^*r_M) + \\ & \quad  ~~~~~~~~~~~~~~~~~~~~ \vdots   \\
& \quad
(\bar{w}_D^*s_{1,D}^*r_1 + \bar{w}_D^*s_{2,M}^*r_2 + \ldots \bar{w}_D^*s_{M,D}^*r_M) \\
 & = \bar{\bf w}^H(i) {\bf S}_D^H(i) {\bf r}(i) = \bar{\bf w}^H(i) \bar{\bf
 r}(i).
\end{split}
\end{equation}

The MMSE expressions for the filters ${\bf S}_D(i)$ and $\bar{\bf
w}(i)$ can be computed through the following optimization problem:
\begin{equation}
\begin{split}
J & = E\big[ | d(i) - \bar{\bf w}^{H}(i){\bf S}_D^H(i){\bf
r}(i)|^2 \big] \\
& = E\big[ | d(i) - \bar{\bf w}^{H}(i)\bar{\bf r}(i)|^2 \big].
\end{split}
\end{equation}
By fixing the projection ${\bf S}_D(i)$ and minimizing (7) with
respect to $\bar{\bf w}(i)$, the reduced-rank filter weight vector
becomes
\begin{equation}
\bar{\bf w}(i) = \bar{\bf R}^{-1}(i) \bar{\bf p}(i),
\end{equation}
where $\bar{\bf R}(i) = E[{\bf S}_D^H(i){\bf r}(i){\bf r}^H(i)
{\bf S}_D(i) ] =E[\bar{\bf r}(i) \bar{\bf r}^{H}(i)]$, $\bar{\bf
p}(i) = E[d^{*}(i){\bf S}_D^H(i){\bf r}(i)] = E[d^{*}(i)\bar{\bf
r}(i)]$. We proceed with the proposed joint optimization  by
fixing $\bar{\bf w}(i)$ and minimizing (7) with respect to ${\bf
S}_D(i)$. We then arrive at the following expression for the
projection operator
\begin{equation} {\bf S}_D(i) = {\bf R}^{-1}(i) {\bf P}_D(i) {\bf
R}_{w}(i),
\end{equation}
where  ${\bf R}(i) = E[{\bf r}(i){\bf r}^{H}(i)]$, ${\bf P}_D(i) =
E[d^{*}(i){\bf r}(i){\bf w}^H(i)]$ and ${\bf R}_w(i) = E[{\bf
w}(i){\bf w}^{H}(i)]$. The associated MMSE is
\begin{equation}
J_{MMSE} = \sigma^{2}_{d} - \bar{\bf p}^{H}(i) \bar{\bf R}^{-1}(i)
\bar{\bf p}(i),
\end{equation}
where $\sigma^{2}_{d}=E[|d(i)|^{2}]$. Note that the filter
expressions in (8) and (9) are not closed-form solutions for
$\bar{\bf w}(i)$ and ${\bf S}_D(i)$ since (8) is a function of
${\bf S}_D(i)$ and (9) depends on $\bar{\bf w}(i)$ and thus it is
necessary to iterate (8) and (9) with an initial guess to obtain a
solution. The MWF \cite{gold&reed} employs the operator ${\bf S}_D
= \big[{\bf p}~{\bf R}{\bf p}~\ldots~ {\bf R}^{D-1}{\bf p}\big]$
that projects the data onto the Krylov subspace. Unlike the MWF
approach, the new scheme provides an iterative exchange of
information between the reduced-rank filter and the projection
matrix and leads to a much simpler adaptive implementation than
the MWF. The projection matrix reduces the dimension of the input
data, whereas the reduced-rank filter attempts to estimate the
desired signal. The key strategy lies in the joint optimization of
the filters. The rank $D$ must be set by the designer in order to
ensure appropriate performance. In the next section, we seek
iterative solutions via adaptive LMS algorithms.

\section{Adaptive LMS Implementation of the Proposed Reduced-Rank Scheme}
\label{sec:typestyle}

In this section we describe an adaptive implementation and detail
the computational complexity in terms of arithmetic operations of
the proposed reduced-rank scheme.

\subsection{Adaptive Algorithms}

Let us consider the MSE cost function
\begin{equation}
J  = E\big[ | d(i) - x(i)|^2 \big] = E\big[ | d(i) - \bar{\bf
w}^{H}(i){\bf S}_D^H(i){\bf r}(i)|^2 \big].
\end{equation}
By computing the gradient terms of (11) with respect to $\bar{\bf
w}(i)$ and ${\bf S}_D(i)$, and using the instantaneous values of
these gradients, one can devise jointly optimized LMS algorithms
for parameter estimation. Let us first describe the computation of
the gradients of (11) with respect to $\bar{\bf w}(i)$ and ${\bf
S}_D(i)$:
\begin{equation}
\begin{split}
\nabla_{\bar{\bf w}(i)} J = \frac{\partial J}{\partial \bar{\bf
w}^*(i)} & = -\big(d(i) - \bar{\bf w}^{H}(i){\bf S}_D^H(i){\bf
r}(i)\big)^* {\bf S}_D^H (i){\bf r}(i) \\
& = -e^*(i){\bf S}_D^H(i){\bf r}(i) = - e^*(i)\bar{\bf r}(i),
\end{split}
\end{equation}
\begin{equation}
\begin{split}
\nabla_{{\bf S}_D(i)} J = \frac{\partial J}{\partial {\bf
S}_D^*(i)} & = -\big(d(i) - \bar{\bf w}^{H}(i){\bf S}_D^H(i){\bf
r}(i)\big)^*
{\bf r}(i)\bar{\bf w}^H(i) \\
& = - e^*(i){\bf r}(i)\bar{\bf w}^H(i).
\end{split}
\end{equation}
By using the gradient rules $\bar{\bf w}(i+1) = \bar{\bf w}(i) -
\mu \nabla_{\bar{\bf w}(i)} J$ and ${\bf S}_D(i+1) = {\bf S}_D(i)
- \eta \nabla_{{\bf S}_D(i)} J$, where $\mu$ and $\eta$ are the
step sizes, the proposed jointly optimized and iterative LMS
algorithms for reduced-rank parameter estimation are
\begin{equation}
\bar{\bf w}(i+1) = \bar{\bf w}(i) + \mu e^*(i)\bar{\bf r}(i),
\end{equation}
\begin{equation}
{\bf S}_D(i+1) = {\bf S}_D(i) + \eta e^*(i){\bf r}(i)\bar{\bf
w}^H(i).
\end{equation}
The LMS algorithms described in (14)-(15) have a complexity
$O(DM)$. In our studies, we verified a performance significantly
superior to full-rank estimation algorithms and that there is no
local minima in the optimization procedure. The proposed scheme
and algorithms trade-off a full-rank LMS adaptive filter against
$D$ full-rank adaptive filters as the projection matrix ${\bf
S}_D(i)$ and one reduced-rank adaptive filter $\bar{\bf w}(i)$
operating simultaneously and exchanging information.

\subsection{Computational Complexity}

Here, we provide the computational complexity in terms of
additions and multiplications of the proposed schemes with LMS
algorithms and other existing algorithms, namely the Full-rank LMS
and the LMS version of the MWF, as shown in Table 1. The MWF has a
complexity $O(D \bar{M}^{2})$, where the variable dimension of the
vectors $\bar{M} = M - d$ varies according to the the rank $d = 1,
\ldots, D$. The proposed scheme is much simpler than the MWF and
slightly more complex than the Full-rank (for $D << M$, as will be
explained later).

{
\begin{table}[h]
\centering%
\caption{\small Computational complexity of LMS algorithms.} {
\begin{tabular}{ccc}
\hline \rule{0cm}{2.0ex}&  \multicolumn{2}{c}{\small Number of
operations per symbol } \\ \cline{2-3}
{\small Algorithm} & {\small Additions} & {\small Multiplications} \\
\hline
\emph{\small \bf Full-rank} & {\footnotesize $2M$} & {\footnotesize $2M+1$}  \\
\emph{\small \bf Proposed}   & {\footnotesize $2DM+D$} & {\footnotesize $3DM+D+2$}  \\
\emph{\small \bf MWF} & {\footnotesize $D(2\bar{M}^{2} -3\bar{M} + 1)$} & {\footnotesize $D(2\bar{M}^{2} +5\bar{M} + 7)$}  \\
\hline
\end{tabular}
}
\end{table}
}

\section{Simulations}
\label{sec:print}

In this section we analyze the proposed reduced-rank scheme and
algorithms in a linear CDMA interference suppression application.
Note that non-linear techniques
\cite{delamare_mber,delamare_itic,delamaretc} are also possible. We
consider the uplink of a symbol synchronous BPSK DS-CDMA system with
$K$ users, $N$ chips per symbol and $L$ propagation paths. Assuming
that the channel is constant during each symbol interval and the
randomly generated spreading codes are repeated from symbol to
symbol, the received signal after filtering by a chip-pulse matched
filter and sampled at chip rate yields the $M$-dimensional received
vector
\begin{equation}
{\bf r}(i)  = \sum_{k=1}^{K} {\bf H}_{k}(i) A_{k} {\bf C}_{k}{\bf
b}_{k}(i)  + {\bf n}(i),
\end{equation}
where $M=N+L-1$, ${\bf n}(i) = [n_{1}(i) ~\ldots~n_{M}(i)]^{T}$ is
the complex Gaussian noise vector with $E[{\bf n}(i){\bf
n}^{H}(i)] = \sigma^{2}{\bf I}$, the symbol vector is ${\bf
b}_{k}(i) = [b_{k}(i+L_{s}-1)~\ldots ~ b_{k}(i)~\ldots ~
b_{k}(i-L_{s}+1)]^{T}$, the amplitude of user $k$ is $A_{k}$,
$L_{s}$ is the intersymbol interference span, the $((2L_{s}-1)
\cdot N)\times (2L_{s}-1)$ block diagonal matrix ${\bf C}_{k}$ is
formed with $N$-chips shifted versions of the signature  ${\bf
s}_{k} = [a_{k}(1) \ldots a_{k}(N)]^{T}$ of user $k$ and the $
M~\times (2\cdot L_{s}-1) \cdot N$ convolution matrix ${\bf
H}_k(i)$ is constructed with shifted versions of the $L\times 1$
channel vector ${\bf h}_k(i) = [{h}_{k,0}(i) ~\ldots ~
{h}_{k,L_{p}-1}(i)]^T$ on each column and zeros elsewhere. For all
simulations, we assume $L=8$ as an upper bound, use $3$-path
channels with relative powers given by $0$, $-3$ and $-6$ dB,
where in each run the spacing between paths is obtained from a
discrete uniform random variable between $1$ and $2$ chips and
average the experiments over $100$ runs. The system has a power
distribution amongst the users for each run that follows a
log-normal distribution with associated standard deviation $1.5$
dB.

We compare the proposed reduced-rank scheme with the full-rank
\cite{diniz} and the MWF \cite{goldstein} implementation for the
design of linear receivers, where the reduced-rank filter
$\bar{\bf w}(i)$ with $D$ coefficients provides an estimate of the
desired symbol for the desired used (user $1$ in all experiments)
as given by
\begin{equation}
 \hat{b}(i) = \textrm{sgn}\Big(\Re\Big[\bar{\bf w}^{H}(i)\bar{\bf
 r}(i)\Big]\Big)=  \textrm{sgn}\Big(\Re\Big[{ x}(i)\Big]\Big)
\end{equation}
where $\Re(\cdot)$ selects the real part, $\textrm{sgn}(\cdot)$ is
the signum function.

We first consider the tuning of the rank $D$ with optimized step
sizes for all schemes, as shown in Fig. 2. The results indicate
that the best rank for the proposed scheme is $D=3$ (which will be
used in the remaining experiments) and it is very close to the
MMSE. Our studies with systems with different processing gains
show that $D$ is invariant to the system size, which brings
considerable computational savings.

\begin{figure}[!htb]
\begin{center}
\def\epsfsize#1#2{0.95\columnwidth}
\epsfbox{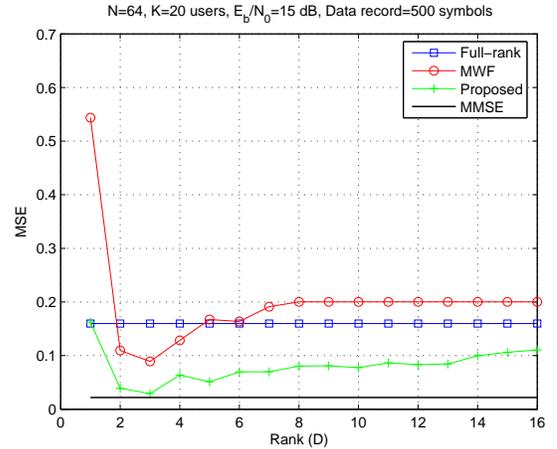} \caption{ MSE performance versus rank (D).}
\end{center}
\end{figure}

We show an experiment in Fig. 3 where the adaptive filters are set
to converge to the same MSE. The LMS-type version of the MWF is
known to have problems in these situations since it does not
tridiagonalize its covariance matrix \cite{goldstein} and thus is
unable to approach the MMSE. The curves show an excellent
performance for the proposed scheme and algorithms, that converge
much faster than the full-rank filter and approaches the MMSE
performance.

\begin{figure}[!htb]
\begin{center}
\def\epsfsize#1#2{0.95\columnwidth}
\epsfbox{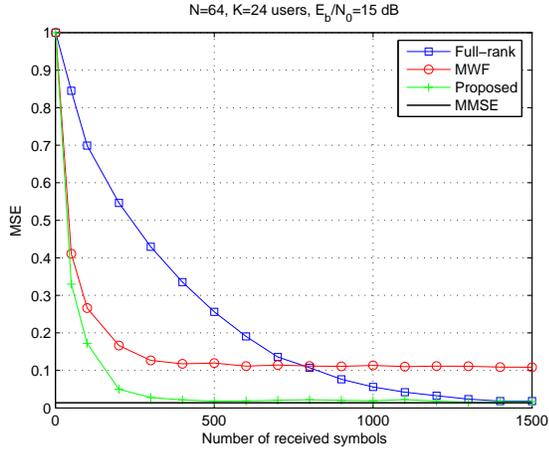} \caption{ MSE performance versus number of
received symbols.}
\end{center}
\end{figure}

The BER convergence performance is shown in Fig. 4 for a mobile
communications scenario. The channel coefficients are obtained
with Clarke´s model \cite{rappa} and the adaptive filters are
trained with $500$ symbols and then switch to decision-directed
mode. The results show that the proposed scheme has a
significantly better performance than the existing approaches and
is able to adequately track the desired signal.

\begin{figure}[!htb]
\begin{center}
\def\epsfsize#1#2{1\columnwidth}
\epsfbox{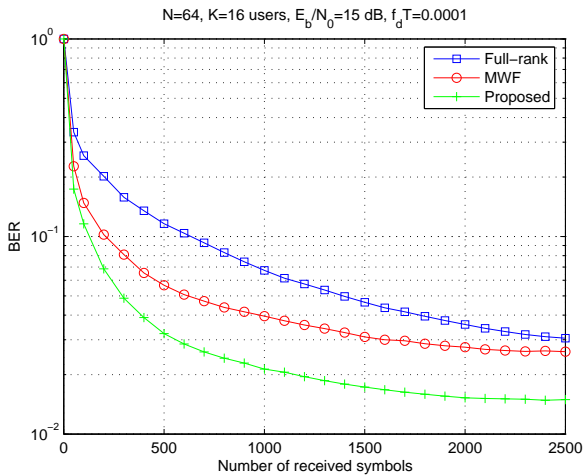} \caption{ BER performance versus number of
received symbols.}
\end{center}
\end{figure}

\section{Conclusions}
\label{sec:conc}

We proposed a novel MMSE reduced-rank scheme based on the joint
and iterative optimization of a projection matrix and a
reduced-rank filter and a low complexity adaptive implementation
using LMS algorithms. The results for CDMA interference
suppression show a performance significantly better than existing
schemes and close to the optimum MMSE.

\end{document}